\documentclass[sigconf, nonacm]{acmart}

\AtBeginDocument{%
  \providecommand\BibTeX{{%
    \normalfont B\kern-0.5em{\scshape i\kern-0.25em b}\kern-0.8em\TeX}}}

\usepackage{dirtytalk}
\usepackage{booktabs}
\usepackage{ragged2e}

\usepackage{colortbl}

\definecolor{maroon}{cmyk}{0,0.87,0.68,0.32}

\begin{document}

\title[Creativity, Generative AI, and Software Development: A Research Agenda]{Creativity, Generative AI, and Software Development:\\A Research Agenda}

\author{Victoria Jackson}
\affiliation{%
  \institution{University of California, Irvine}
  \city{Irvine}
  \country{USA}}
\email{vfjackso@uci.edu}

\author{Bogdan Vasilescu}
\affiliation{%
    \institution{Carnegie Mellon University}
    \city{Pittsburgh}
    \country{USA}}
\email{vasilescu@cmu.edu}

\author{Daniel Russo}
\email{daniel.russo@cs.aau.dk}
\orcid{0000-0001-7253-101X}
\affiliation{%
  \institution{Aalborg University}
  \streetaddress{A.C. Meyers Vaenge 15, 2450}
  \city{Copenhagen}
  \country{Denmark}}

\author{Paul Ralph}
\affiliation{%
  \institution{Dalhousie University}
  \city{Halifax}
  \country{Canada}}
\email{paulralph@dal.ca}

\author{Maliheh Izadi}
\affiliation{%
  \institution{Delft University of Technology}
  \city{Delft}
  \country{Netherlnads}}
\email{m.izadi@tudelft.nl}

\author{Rafael Prikladnicki}
\affiliation{%
  \institution{PUCRS University}
  \city{Porto Alegre}
  \country{Brazil}}
\email{rafaelp@pucrs.br}

\author{Sarah D'Angelo}
\affiliation{%
  \institution{Google}
  \city{Auckland}
  \country{New Zealand}}
\email{sdangelo@google.com}

\author{Sarah Inman}
\affiliation{%
  \institution{Google}
  \city{Seattle}
  \country{USA}}
\email{icsarah@google.com}

\author{Anielle Lisboa}
\affiliation{%
  \institution{PUCRS University}
  \city{Porto Alegre}
  \country{Brazil}}
\email{anielle.lisboa@edu.pucrs.br}

\author{Andr\'e van der Hoek}
\affiliation{%
  \institution{University of California, Irvine}
  \city{Irvine}
  \country{USA}}
\email{andre@ics.uci.edu}

\renewcommand{\shortauthors}{Jackson, Vasilescu, Russo, Ralph, Izadi, Prikladnicki, D'Angelo, Inman, Lisboa, van der Hoek}

\begin{abstract}
Creativity has always been considered a major differentiator to separate the good from the great, and we believe the importance of creativity for software development will only increase as GenAI becomes embedded in developer tool-chains and working practices. This paper uses the McLuhan tetrad alongside scenarios of how GenAI may disrupt software development more broadly, to identify potential impacts GenAI may have on creativity within software development. The impacts are discussed along with a future research agenda comprising six connected themes that consider how individual capabilities, team capabilities, the product, unintended consequences, society, and human aspects
can be affected.

\end{abstract}



\received{29 March 2024}

\maketitle

\section{Introduction}
Creativity has always been important to software development \cite{glass1994software}. Creativity helps teams address the more mundane software tasks that arise everyday as well as the infrequent tasks leading to major product advances. For example, everyday creativity (Little-c) \cite{amabile_pursuit_2017}, in the form of creative problem solving, helps resolve the challenges that inevitably arise in commonplace tasks needed to build software (e.g., debugging, refactoring, just coding a new feature). The more infrequent \say{big-bang} Big-C creativity \cite{kaufman_beyond_2009} is needed when major innovations are required (e.g., perhaps in re-architecting an application to incorporate the latest Artificial Intelligence (AI) techniques or identifying new product features that will delight customers). To flourish, both forms of creativity rely on the interplay of talented individuals, collaborative teams, supportive environments, and use of appropriate techniques (and supportive tools) \cite{amabile1988model} such as whiteboarding \cite{groeneveld2021exploring}, ideation, and hackathons \cite{granados_how_2019}). 

While understudied \cite{amin2018snapshot}, we are at a watershed moment when it comes to creativity in software development, since the emergence of Generative AI (GenAI) could have a major impact on individuals, the products they build, and their employers. We already see that GenAI can help developers\footnote{We use the term developer to cover all roles in a software team and not just someone who writes code.} be creative by generating ideas \cite{bockeler_generative_nodate}, which can thereby influence the resulting products. However, if GenAI becomes as powerful as some people believe in terms of taking the \say{rote} out of software development \cite{freethink_2023}, it could be argued that all that is left in software development is creativity and that companies will increasingly view creativity as being a competitive advantage and a powerful differentiator from rival products---more so than today even. 


Therefore, although a significant portion of research, studies, and discussions in the media have concentrated on GenAI and programmer productivity (e.g., \cite{10.1145/3633453, bird2023taking}, we argue that understanding the short-term, medium-term, and long-term impacts and connections between creativity and GenAI in software development warrants equal, if not greater, emphasis. As a first step to drawing attention to this issue, 
this paper makes two contributions:
\begin{enumerate}
    \item With the help of the Marshall McLuhan tetrad \cite{mcluhan1977laws} along with a range 
    of potential scenarios that may play out as to how GenAI will shape software development, we contribute an in-depth set of potential impacts GenAI may have on the role of creativity within software development. The impacts are organized along four components of creativity: the Person, the Product, the Process, and the Press (the 4P framework \cite{rhodes1961analysis}). Using the 4P framework allows us to provide both breadth and depth to our analysis.
    \item Derived from the potential impacts, we propose a research agenda, comprising of six distinct themes that build upon one another. 
    These themes, within the context of impact of GenAI on creativity in software development, focus on tangible and immediate effects such as individual capabilities, team capabilities, and products, as well as indirect effects concerning unintended consequences, society, and human aspects. Overall, our work calls for future research studying these themes alongside the ongoing evolution of GenAI, to allow potential interventions for preventing harmful impacts as well as amplifying positive impacts.
\end{enumerate}    

\section{Definitions}
While many definitions of creativity exist \cite{runco_standard_2012}, we adopt Boden's definition \say{Creativity is the ability to come up with ideas or artifacts that are new, surprising and valuable} \cite{boden2004creative}. This definition talks to the act of being creative (the process) and also the outcome (the product), both of which are relevant to software development. 

To explain creativity, various theoretical frameworks have been proposed over the years (e.g., the Systems Model of Creativity \cite{csikszentmihalyi_society_2014}, the Componential Model of Creativity \cite{amabile1988model}). In this paper, we utilize the 4P framework \cite{rhodes1961analysis}. This framework considers that creativity can be explained by four components: the \textbf{Person}, the \textbf{Product}, the \textbf{Process}, and the \textbf{Press} (environment). Person refers to the creative individual and consists of their personality, behaviors, and skills that contribute to their creativity. Product is the outcome of the creative process -- in the case of software development, this could be an artifact such as code or a document or the actual end-user product. Process is the methods and techniques used to generate new ideas, such as brainstorming. Finally, Press acknowledges the impact of the environment (e.g., social, cultural, economic, physical) on creativity. Adopting this framework allows us to explore the potential impact of GenAI on software development from a comprehensive, structured set of dimensions.

\section{Related Work}
We briefly review related work on creativity within software development and the use of Gen AI within software development.

Generally, creativity within software development is an understudied topic \cite{mohanani2017perceptions,inman2024developer}. For Person creativity, only a few studies exist. One finds that personality traits can predict a programmer's creativity \cite{amin2020impact} and another that having overly templated requirements can result in fixation \cite{mohanani2021templated} when designing software. Product creativity has been little examined beyond a few studies on Open-Source Software (OSS), with some claiming they are more creative than commercial software \cite{o1999lessons, kidane2007correlating} and another framing innovation in OSS projects as novel reuse of existing libraries \cite{fang2024innovation}. Within Process creativity, much research has looked into requirements engineering \cite{amin2018snapshot}, such as techniques to encourage creative thinking \cite{maiden_provoking_2004} in identifying requirements. Whiteboarding and brainstorming are popular creativity techniques used by developers \cite{groeneveld2021exploring}. These techniques, alongside other techniques such as mob programming \cite{staahl2021mob}, are also suited for fostering creativity in hybrid teams \cite{jackson2022team}. The environment (Press) can affect an individual's creativity. Working in a psychologically-safe climate \cite{edmondson2014psychological} can aid creativity \cite{zadow_psychosocial_2023}, as can having a supportive manager who encourages new ideas \cite{da_silva_how_2016}. Having high levels of team collaboration, however, can hinder creativity \cite{hoegl_creativity_2007}. Distance can lead to reduced creativity since distributed designers spend less time exploring the problem space than co-located designers \cite{jolak2023design}.

The adoption of Large Language Models (LLMs), a form of GenAI, for software development has increased dramatically in the last few years. Two recent literature reviews \cite{fan2023large, hou2024large} 
find the majority of research has focused on code-generation tasks followed by testing. Neither review mentions studies that directly target creativity, although it is indirectly hinted at when discussing potential research into hallucinations and whether the hallucinations could be harnessed in some way, e.g., identifying new features \cite{fan2023large}. Both papers also include research into the use of LLMs for tasks that developers typically consider as creative (e.g., bug fixing), showing the potential for LLMs to disrupt creative work. 

To summarize, to the best of our knowledge, there is no research that directly examines the impact of GenAI on creativity in software development. This is perhaps unsurprising given the lack of research on creativity in software development more generally. Today, however, it is more important than ever to truly engage in research on creativity, GenAI, and software development. Such is the purpose of this paper: to raise awareness of the issue and identify important directions for future research.

\section{Exploring the Potential Impact}
We took a two-phased approach to exploring the potential impact of GenAI on creativity in software development. Firstly, we considered future scenarios of how much GenAI may impact creativity from a human perspective. Secondly, we adopted parts of the Disruptive Research Playbook \cite{storey2024disruptive} (from now on referred to as `playbook'), which was specifically designed for identifying socially relevant software development research questions when studying disruptive technologies (such as GenAI). 

First, much has been said about how GenAI will impact software development, from the mundane to the fantastical. Some predict, for instance, that GenAI will simply be another tool in the tool belt of software developers, amplifying their ability to perform work but not replacing their jobs~\cite{10176168}.  Others compare the current disruptive impact of GenAI on many professions (including software developers) to that of 19th-century weavers facing changes due to the invention of automated looms \cite{merchant_blood_2023}, or predict dire job losses in the face of increased workplace automation \cite{kochhar_which_2023}.  Any research agenda should include the possible spectrum of impact that may transpire, which is why we drew up a series of brief, high-level scenarios that together cover this spectrum.

Second, the playbook is built upon the McLuhan tetrad \cite{mcluhan1977laws} to frame the research landscape. The tetrad is designed to help consider the potential impact of a new technology by posing four questions about what the technology: (i) \textbf{enhances}, (ii) makes \textbf{obsolete}, (iii) \textbf{retrieves} from obsolescence, and (iv) \textbf{reverses} into when pushed to extremes. The answers to these questions can then be used to choose specific phenomena to study, identify potential research questions, and determine research strategies best suited to answering them. Using the variety of scenarios, we brainstormed impacts to populate a tetrad for each P of the 4P framework of creativity. In this context, it is important to note that the tetrad merely provides a framework for identifying future possibilities, with a fair degree of imagination and brainstorming required to populate the tetrad (or in our case, tetrads). Different groups, thus, may produce a set of questions and impacts different from the authors of this paper. Indeed, we do not consider our tetrads exhaustive and encourage others to further augment our ideas.


\subsection{Future Scenarios}

It is too soon to know fully where we eventually will end up as GenAI continues to advance and become more integrated into the daily routines of software developers. Based on early observations, we do know GenAI is likely to affect how developers practice creativity \cite{bockeler_generative_nodate}. However, projections of what GenAI can and cannot achieve in terms of creativity support differ significantly. From a human perspective, the two extreme scenarios are the \say{Optimistic} versus \say{Pessimistic} future. In the \say{Optimistic} scenario, creativity increases to create ever more exciting products with developers actively involved. This is because developers have more time for creative thinking, as they offload the more mundane work of software development to the GenAI. The additional time provides developers with ample slack time for deep thought, and to use GenAI to brainstorm, discuss potential ideas, and involve it as a fellow creative partner. The \say{pessimistic} scenario is that GenAI advances to the point where it can do almost all the work of software developers and so the role of humans in development work is greatly diminished. There is no need for creativity from humans; instead, GenAI is the creative force. Amazing new products result from the creativity of GenAI, yet the human is no longer part of the process. 

In between these two extremes are a great many other scenarios.  One is where GenAI helps in programming and design work, but product feature work remains the sole responsibility of the developer. Another might be that GenAI can become a useful agent that participates in all aspects of the development process like a true team member, but one such agent is sufficient to complement a team in their work because it learns how the team works and what kinds of problems it tackles. In yet another scenario, multiple GenAIs could be working together autonomously to identify and explore a set of features that they deliver to a product team for further consideration.

While space prohibits us from detailing all scenarios and particularly from including how they shape the human experience in each case,
we used them to drive our brainstorming in the next section. We note that the scenarios are crafted from the perspective of the capabilities of GenAI, but clearly impact developers and their teams in many different ways, most directly in terms of what work they perform how, but indirectly also in their feelings, sense of agency, ability to hone their craft, and well-being. These human factors must equally be part of the research agenda and the tetrads that we present in the next section---as based on considering the scenarios and consequences of these scenarios---integrally consider them.

\begin{table*}
 \centering
 \renewcommand{\arraystretch}{1.25} 
 
\begin{tabular}{p{0.47\linewidth} | p{0.47\linewidth}}
\hline 
\rowcolor{lightgray}
\textbf{ENHANCES} 
\par
\textit{How does GenAI enhance or amplify an individual's creativity?} &
\textbf{OBSOLESCES}
\par
\textit{Which factors considered relevant to an individual's creativity does GenAI make obsolete?}\\
\hline 

\par
\textit{Idea Generation} to help provide a starting point, avoid design fixation, and nudge them in a different direction. \par
\textit{Cross-Domain Inspiration} by exposing a developer to ideas and solutions from other fields. \par
\textit{Checks and Balances} by generating a checklist of things to think about as they work. \par
\textit{Impersonation of downstream stakeholders} e.g., a tester to identify edge cases. \par
\textit{Assessment} of an individual's competing designs to architectural fitness functions. \par
 &  
 \par
 \textit{Differences in personality traits} With GenAI, anyone can be creative irrespective of whether they have a trait commonly associated with creativity; e.g., Openness.
 \par
 \textit{Deep Thinking} Can offload the thought process to GenAI so time for deep thinking is no longer required.
 \par
 \textit{Reflection} Long considered important to breaking through challenging problems, reflection on the problem or solutions is no longer required.
 \par
 \textit{Expertise} is no longer required as knowledge is provided by GenAI.
 \par
 \textit{Mentorship} by other developers is diminished as GenAI provides personalized feedback and learning opportunities.
 \\  
 \hline

\rowcolor{lightgray} 
\textbf{RETRIEVES} 
\par 
\textit{What factors relevant to an individual's creativity does GenAI retrieve that had been made obsolete earlier?}
& \textbf{REVERSES INTO} 
\par
\textit{What does GenAI do to an individual's creativity when GenAI is pushed to extremes or overused?}
\\
\hline
 
\par
 \textit{Sketching} With more time and to balance interactions with GenAI, the practice of manually sketching designs and random doodling may make a comeback and stimulate creativity.
 \par
 \textit{Verification} by reading and interpreting the ideas and solutions generated by GenAI to ensure they're valid will be important.
 \par
 \textit{Creative Work} Today, creativity often takes a backseat to productivity. With GenAI taking the burden of writing code, developers will have more time to engage in fun, creative work.
 &
 \textit{No second guessing} As developers learn to trust and rely on the GenAI, they stop stepping back to consider the solution and no longer push their own thinking.
 \par
 \textit{Loss of consideration of alternatives} A human tendency is to be biased towards the ideas near the top of a list which may lead to insufficient exploration of alternatives to those generated by GenAI.
 \par
 \textit{Loss of problem-solving skills} As GenAI takes over more problem-solving tasks, developers lose their critical thinking and problem-solving skills so unable to address situations when GenAI fails. \\
\end{tabular}
  \caption{McLuhan's tetrad that considers the impact of GenAI on the creativity of a software developer (Person)}
    \label{tab:person_tetrad}
\end{table*}

\begin{table*}
 \centering
\begin{tabular}{p{0.47\linewidth} | p{0.47\linewidth}}
\hline
\rowcolor{lightgray} 
\textbf{ENHANCES} 
\par 
\textit{How does GenAI enhance creative outcomes?}& \textbf{OBSOLESCES} 
\par
\textit{Which creative outcomes become obsolete due to GenAI?}
\\ 
\hline
\par
\textit{Continuous Enhancement} With continuous deployment as the backdrop, the use of GenAI speeds up the stream of new, innovative, and valuable product features. \par
\textit{Customized User Experiences} GenAI can enable the creation of highly personalized and adaptive user experiences, making products more appealing and useful to individual users. \par
\textit{Taking Stock} GenAI analyzes market trends, user feedback, and product performance to help companies identify product opportunities and enhancements. \par
\textit{Making the Impossible Possible} GenAI helps teams develop software that cannot be produced today and benefits society (e.g., universal healthcare records). \par
\textit{If Only We Had} GenAI identifies entirely new classes of products that humans never realized were needed, essentially promoting radical innovation.\par
 &  
 \par
 \textit{Routine Solutions} Typical solutions will become a thing of the past because GenAI always comes up with tailored solutions that fit the situation just right.
 \par
 \textit{Frameworks} Since every application will be generated quickly from the ground-up, general-purpose frameworks are no longer required.
 \par
 \textit{Intermediate Design Artifacts} With GenAI, typical intermediate design artifacts such as concept sketches, wire-frames, UML diagrams, and the like may no longer be needed.
  \par
 \textit{Art} The art that goes into software, such as iconography, color schemes, and GUI layouts, is potentially all generated by the AI, no longer needing to be designed separately.
   \par
 \textit{Specialty Artifacts} Because GenAI inherently knows about everything, specialty artifacts that once were considered the domain of creatives (e.g., music and level design in computer games, clever database design for an enterprise system) are simply generated, and no longer creative human outputs. 
 \\
\hline
\rowcolor{lightgray} 
\textbf{RETRIEVES} 
\par 
\textit{What previous creative outcomes could GenAI bring back to the foreground?}
& \textbf{REVERSES INTO} 
\par
\textit{If overtly relied upon, how could GenAI disrupt creative outcomes?}
\\
\hline 
\par
 \textit{Patterns and Styles} While still considered important overall, thousands of patterns and styles exist that are much less frequently accessed and leveraged than the handful everyone knows. These can now be readily accessed through GenAI.
 \par
 \textit{Old Design} Old designs embedded in old software that may have well been forgotten could work very well for newer situations, whether outright or redressed in a new interface or language.
 \par
 \textit{Analog Design Principles} There might be a renewed interest in analog design principles and their integration into digital products, driven by GenAI’s ability to synthesize and apply wide-ranging design philosophies.
 &
  \par
 \textit{Echo Chamber} The more new products resemble common products that GenAI ‘understands’, the more GenAI’s suggestions could become self-reinforcing, thereby limiting exposure to truly novel ideas.
  \par
 \textit{Homogeneous Products} An over-reliance on GenAI with different developers all choosing the same recommended ideas will reduce the diversity of products in the market.
  \par
 \textit{Overly Creative Products} GenAI’s suggestions can be too creative for a given situation, to the point where adopting them would lead to software that is too complex or too different, where a routine product would have sufficed.
   \par
 \textit{Biased Products} GenAI invariably contains various biases. As a result, products could be exclusionary to certain users or become too woke or too general when GenAI developers overcompensate.
   \par
 \textit{Harmful Products} The use of GenAI could lead to a multiplicity of products that contain harmful features (e.g., dark UI patterns to trick users into undesirable actions, buggy code with security risks that inadvertently enter via GenAI).
\end{tabular}
  \caption{McLuhan's tetrad that considers the impact of GenAI on Product creativity}
    \label{tab:product_tetrad}
\end{table*}

\begin{table*}
 \centering
\begin{tabular}{p{0.47\linewidth} | p{0.47\linewidth}}
\hline
\rowcolor{lightgray} 
\textbf{ENHANCES} 
\par 
\textit{How does GenAI enhance creative processes?}& \textbf{OBSOLESCES} 
\par
\textit{Which creative processes become obsolete due to GenAI?}
\\ 
\hline
\par
\textit{Mundane Activities} Important activities that feed into creativity (e.g., user research, competitive analysis) are performed by GenAI, relieving the developer from having to engage in these.
 \par
 \textit{Rapid Prototyping} GenAI greatly speeds up the process of going from rough ideas to prototypes, experimentation, reflection, and iteration because it automates many mundane tasks in the process; as a result, time is freed up for the team to consider many more ideas. 
\par
 \textit{Enhanced Collaboration} GenAI helps bridge diverse team members and diverse teams by synthesizing inputs, explaining different perspectives, helping overcome terminology barriers, and suggesting integrative solutions that may not be obvious. 
 \par
 \textit{Creativity Needed} GenAI understands when and what kind of creativity is needed at what points in the development process; it suggests the best teams and creates tasks for the teams to address the emergent needs.
\par
 \textit{Moderation} Brainstorming and other creative activities are no longer led by a human, but GenAI orchestrates these activities entirely, continuously steering participants in the right direction.
\par
 \textit{Autonomous Creativity} Multiple GenAIs, under the leadership of an overarching GenAI, participate in creative idea generation and synthesis, taking the place of humans entirely or near-entirely.
& 
\par
 \textit{In-person Brainstorming} Structured or unstructured idea generation is no longer needed since GenAI takes the place of these kinds of human sessions.
\par
 \textit{Siloed Specializations} Divisions between specialized roles in organizations and creative processes may blur as GenAI tools enable individuals to contribute across a broader range of tasks.
\par
 \textit{Linear Processes} The typically linear stages of design first with implementation next might be replaced by more fluid, dynamic GenAI processes that blur the two in real-time.
\par
 \textit{Design Techniques} Techniques such as user studies, focus groups, cognitive walkthroughs, and others are no longer needed since GenAI takes their place.
 \par
 \textit{Hiring IDEO and other such companies} With GenAI, specialist design, and visioning companies are no longer needed to develop innovative approaches to complex problems.
 \\  
 \hline
\rowcolor{lightgray} 
\textbf{RETRIEVES} 
\par 
\textit{What previous creative processes could GenAI bring back to the foreground?}
& \textbf{REVERSES INTO} 
\par
\textit{If overtly relied upon, how could GenAI disrupt creative processes?}
\\
\hline 
\par
 \textit{Pair Programming} Programming in pairs or even mobs is a vehicle for creative problem-solving. Pair or mob programming happens today, but could become much more important to address the remaining challenging problems GenAI cannot solve.
 \par
 \textit{Idea Generation Techniques} To counter GenAI providing mundane solutions, organizations prioritize and amplify events and techniques for out-of-the-box thinking, including brainstorming, hackathons, design thinking sprints, etc.
\par
 \textit{Analog Creativity Techniques} There might be a resurgence in the use of analog creativity techniques, as teams seek to balance GenAI’s capabilities with tangible, hands-on methods that foster deep thinking and innovation.
\par
 \textit{Human-Centered Design} As GenAI handles more of the technical workload, creative processes may increasingly focus on the human side of design, emphasizing empathy and user experience. Again, this happens today, but could become a major focus for the majority of developers.
\par
 \textit{Manual User Research} To avoid becoming entirely disconnected from the customers, manual user research makes a resurgence to shape how the company or team uses GenAI to shape their products.
\par
 \textit{Studying Other Domains} A source of inspiration is to study other systems, other domains, and even other disciplines. With extra time, and the pressure to come up with novel ideas all the time, developers engage more with adjacent and other systems and fields.
&
\par
 \textit{Loss of Intuitive Decision Making} The nuanced, intuitive aspects of decision making that often pervade how teams ultimately make the right choices might be undermined as teams grow accustomed to deferring to GenAI’s data-driven suggestions that ‘must be right’.
\par
 \textit{Loss of Facility with Creativity Techniques} As more and more decisions rely on GenAI, developers practice creativity less and less, thereby slowly but surely eroding teams’ and organizations’ ability to independently perform creative work.
\par
 \textit{Loss of Trust With One Another} Since developers rarely get to work with others on difficult problems where differences of opinion and arguments arise, they lose their ability to constructively disagree, which is a key ingredient for the emergence of creative solutions.
\par
 \textit{Roteness of the developer role} The GenAI has taken over all creative duties and all that remains to the developer is to fill in rote work that GenAI cannot complete. 
\end{tabular}
  \caption{McLuhan's tetrad that considers the impact of GenAI on the creative Process}
    \label{tab:process_tetrad}
\end{table*}

\begin{table*}
 \centering
\begin{tabular}{p{0.47\linewidth} | p{0.47\linewidth}}
\hline
\rowcolor{lightgray} 
\textbf{ENHANCES} 
\par 
\textit{How does GenAI enhance the environmental conditions that promote creativity?}& \textbf{OBSOLESCES} 
\par
\textit{Which environmental conditions become obsolete due to GenAI?}
\\ 
\hline
\par
\textit{Contextual Information} GenAI listens in on the conversations taking place and continuously shares possibly relevant information in a variety of formats (e.g., smart reminders on devices, projection of requirements, potential design options, etc. on walls). \par
\textit{Environmental Inspiration} GenAI can curate environmental settings (e.g., light, sound, visuals) to inspire creativity based on the task at hand (whether for an individual or a team working together), even dynamically adapting settings based on progress. \par
\textit{Resource Management} By optimizing the allocation and use of resources in a creative workspace, GenAI can ensure that creative teams have what they need, when they need it, without the distractions of managing logistics.\par
\textit{Virtual Effectiveness} GenAI improves virtual workspaces and interactions with immersive tools and features that approach or even surpass the benefits of physical co-location, fostering creativity among distributed and/or hybrid teams.\par
\textit{Voice of Reason} GenAI prevents software that brings harm to humans and other species from being developed and deployed.\par
\textit{Increased Risk Taking} It will be quicker to release software changes, so companies may take more risks as it is quicker to pivot and/or change if there are issues.
 &  
 \par
 \textit{Whiteboards, sticky notes, etc.} Traditional tools used in creative activities may no longer be needed.
 \par
 \textit{Cubicles} Traditional cubicles are no longer needed, as software development has become all about creativity, with no more rote tasks that require individual concentration and focus.
\par
 \textit{Designated Spaces} Fixed layouts and functions of certain spaces (e.g., meeting rooms, lounge spaces) may disappear in favor of spaces that dynamically adjust to facilitate different modes of creative work.
 \par
 \textit{Physical Presence} Requirements for physical presence to promote chance encounters and scheduled in-person creative work may diminish or disappear altogether.
\par
 \textit{Offices} With creativity the sole surviving skill, work can be performed anywhere, with no need for an office, whether at work or at home. The park is a great place to think.
\par
 \textit{Colleagues} With GenAI support, developers can design, create, and deploy any system they can imagine, alleviating the need for colleagues in the creative process, whether informally via typical chance encounters or formally through explicitly working as a team.
 \\  
 \hline
\rowcolor{lightgray} 
\textbf{RETRIEVES} 
\par 
\textit{What environmental conditions could GenAI bring back to the foreground?}
& \textbf{REVERSES INTO} 
\par
\textit{If overly relied upon, how could GenAI disrupt environmental conditions?}
\\
\hline
 \par
 \textit{Physical Movement} As GenAI takes over tasks that traditionally required being seated at a desk, there may be a revival of designing spaces that encourage physical movement and its relationship to creative thinking.
\par
 \textit{Supportive Management} Oftentimes, creativity – especially everyday creativity – is not recognized as part of day-to-day business. Now individuals have the freedom to express their creativity and be recognized for it by management because creativity is a strategic differentiator for the business.
 &
 \par 
\textit{Sabotage} Software developers are so enraged with their jobs being taken, they sabotage GenAI so that it under-performs and underwhelms, eventually becoming obsolete.
 \par
 \textit{No Physical Space} As virtual creative environments become more sophisticated, there is a risk that physical creative spaces are eliminated because their value to and tangible benefits for creativity are underestimated. 
\par
 \textit{Isolation} While enhancing virtual collaboration in creative exercises, an overemphasis on technology could lead to isolation, as individuals may interact more with GenAI than with each other, potentially stifling spontaneous human interactions that spark creativity. 
 \par
 \textit{Reduced Pride} Developers feel they have less agency to be creative, as GenAI does it all for them. They lose motivation, pride, and enjoyment from their work as they can no longer express themselves creatively. They may quit or suffer poor mental well-being.
\par
 \textit{Management Pressure} Greater expectations on individuals with managers pushing even harder for developers and teams to be creative. This pressure is probably not appreciated by all and may lead to serious friction.
 \par
 \textit{Unable to Speak Up} Management places so much trust in the creativity of GenAI that individuals feel unable to speak up in dissent of the approaches GenAI takes.
 \par
  \textit{Creativity What} Because everything is creative, creativity is no longer appreciated as anything special and becomes an undervalued skill. 
\end{tabular}
  \caption{McLuhan's tetrad that considers the impact of GenAI on the creative environment (Press)}
    \label{tab:pres_tetrad}
\end{table*}

\subsection{Tetrads}
To use the McLuhan tetrad (Step 1 of the playbook), a technology must be selected as the source of the disruption. For this paper, it is GenAI. We also limit the phenomena under consideration (Step 2 of the playbook) to solely creativity in software development, rather than other phenomena common to software development also impacted by GenAI (e.g., productivity, code quality).

Combining GenAI, creativity in software development, and the 4P framework of creativity (Person, Product, Process, Press) requires four tetrads; one for each of the Ps. Taking the Person as an example, the tetrad requires answers to the following four questions:
\begin{enumerate}
    \item How does GenAI \textbf{enhance} or amplify an individual's creativity?
    \item Which factors considered relevant to an individual's creativity does GenAI make \textbf{obsolete}?
    \item What factors relevant to an individual's creativity does GenAI \textbf{retrieve} that had been made obsolete earlier?
    \item What does GenAI do to an individual's creativity when GenAI is \textbf{pushed to extremes} or overused?
\end{enumerate}

Collectively, the researchers brainstormed a set of answers across each of the four tetrads. When brainstorming, the researchers referred to the future scenarios discussed earlier and also considered the impact on both little-c everyday creativity and the Big-C creativity common to major innovations, so to cover both the day-to-day problem-solving that drives software development all the time and the more infrequent, high-level visioning of products and their features.  Tables \ref{tab:person_tetrad} - \ref{tab:pres_tetrad} show the four resulting tetrads.
We encourage the reader to carefully study the content of each, and perhaps augment them with their own projections as answers to the questions.

Continuing with the example of the Person component of creativity, Table \ref{tab:person_tetrad} contains many potential impacts of GenAI on the Person, out of which we highlight a few that we consider particularly important.  Note that we do not necessarily predict all of these will happen; rather, by considering the breadth of potential implications, a thoughtful research agenda can be shaped. While creativity could be enhanced by GenAI \textbf{assisting with idea generation} and providing \textbf{cross-domain inspiration}, there is a risk that at the extremes developers rely so much on GenAI for creativity that they \textbf{stop considering alternatives} and \textbf{lose problem-solving skills} critical to software development. The potential also exists for GenAI to \textbf{obsolete factors known to aid creativity such as personality traits and expertise}, as these are no longer relevant when GenAI helps the individual be creative.
GenAI could indeed be a great leveler, bringing all developers to the same creative level.  However, using GenAI will require developers \textbf{to continue to hone their verification skills} to ensure the generated ideas are valid.

Product creativity (Table \ref{tab:product_tetrad}) may be enhanced as the use of GenAI allows the \textbf{rate of delivery of new, innovative features to be increased} alongside \textbf{customized user experiences}. Although there is a risk that \textbf{products become homogenized} due to an over-reliance on GenAI to determine new features as well as a risk that \textbf{harmful products incorporating dark patterns} could proliferate. In using GenAI, \textbf{common frameworks may no longer be required} as all code is generated from the ground-up, but \textbf{interest in analog design principles may be revived} due to GenAI incorporating wide-ranging design philosophies.

Since GenAI is a tool, it is not surprising that the creative process could be impacted in numerous ways (Table \ref{tab:process_tetrad}), such as a creative process that benefits from GenAI \textbf{automating more mundane tasks} or \textbf{GenAI acting as the moderator} in practices such as brainstorming and leading to better outcomes. However, if GenAI is relied upon too much, \textbf{developers lose their own creative abilities}, including intuitive decision making. \textbf{Siloed specializations may become eliminated} as GenAI enables developers to contribute to creative work outside their own area. Moreover, \textbf{the need for design techniques} such as user studies with real people may become obsolete, as GenAI can simulate the people. However, using GenAI \textbf{may further promote tasks used today to stimulate creativity} such as pair programming and idea generation techniques in order to solve problems too complex for GenAI or to provoke solutions that differ from GenAI generated ones.

Ways in which the creative environment (Table \ref{tab:pres_tetrad}) could be impacted include \textbf{contextual information being provided automatically} by GenAI listening in and offering suggestions or, taken to extremes, \textbf{developers losing pride in their work} as GenAI does all of the interesting creative work. \textbf{Colleagues may no longer be needed} as GenAI can do all the support work needed to design, build, and deploy a new system. It may also \textbf{be more important for management to be supportive of developer creativity} as creativity will be a strategic business differentiator.


Taken collectively, the tetrads reveal many potential implications in using GenAI for creativity within software development. Some of these would appear to be more positive for the developer community (e.g., more time for creative, fun work) whereas others clearly exacerbate issues known today (e.g., increased sense of isolation for remote workers as virtual collaboration is no longer needed when one can brainstorm with the GenAI). Not all the implications noted here may come to pass, but collectively they reveal the breadth of considerations that future research must take into account.

\section{Research Agenda}

The four tetrads (Tables \ref{tab:person_tetrad}--\ref{tab:pres_tetrad}) list seventy-six potential impacts of using GenAI for creativity in software development. Each of them could be considered a hypothesis, to be tested by designing a specific research question along with determining the most appropriate research method to explore it (Steps 3 and 4 of the Disruptive Research Playbook \cite{storey2024disruptive}). 
However, not all are as likely to emerge nor are they all as immediate in their impact. Depending on GenAI's continued evolution, some may indeed never come to fruition and some others may take a shape we are not considering yet at this time. Rather than turning each impact into a research question and associated method, we therefore instead organize our research agenda for creativity, GenAI, and software development along six high-level themes: (i) individual capabilities, (ii) team capabilities, (iii) the product, (iv) the unintended consequences, (v) societal impact, and (vi) the human aspects.
Each of these themes will provide varying perspectives on the impact of GenAI on creativity, as they will investigate different units of analysis and collectively provide a holistic view. The latter three themes start to converge on themes that scholars from other domains are concerned with --- the broader impact of GenAI generally, and not just creativity, on all of society.



The outcomes from the numerous research studies will no doubt include insights into both the social and the technical. The ways people work, including new ones, how people feel, and the way they use tools (or need new ones) are just some likely findings to emerge from future studies. Social implications will be keenly understood as the unintended consequences, societal impact, and human aspects are studied. In some ways, these social implications are the most important to understand as they inform researchers and practitioners who design and build GenAI of the potential harms caused by the use of GenAI for creative software work, in much the same way developers are only now grappling with the issues caused by biases in LLMs \cite{10.1145/3582269.3615599}.



\subsection{Individual Capabilities}
Many of the identified impacts across the tetrads imply that creativity of the individual will increase, for two overarching reasons: (1)~GenAI provides creativity support through idea generation, including cross-domain inspiration, and (2) GenAI taking on mundane aspects of software development and even creative work, which provides more time for pondering, doodling, and coming up with a creative solution.
However, it is unclear if the individual will truly be more creative and in what aspects 
of their work
(if at all).
Various natural human responses may lead to less creative exploration (e.g., no second guessing, loss of consideration), and necessary skills may deteriorate over time. At the most fundamental level, then, research is needed to study whether GenAI helps developers be more creative, and if so, for what tasks, how, and whether it persists over time. Given that this can depend on how GenAI is made available to them and is integrated into their day-to-day tool usage, experimentation with new kinds of tools and studying emerging working practices will be equally important.

\subsection{Team Capabilities}
While ``team'' does not appear as a tetrad, there is an implied impact on teams and their ability to be creative across a number of the tetrads, especially in the Process tetrad (Table \ref{tab:process_tetrad}) but also the Press tetrad (Table \ref{tab:pres_tetrad}) as the environment surrounding a team can influence the creativity exhibited. Our concern here goes beyond multiple team members using GenAI individually; rather, the core question is whether GenAI interjected in teamwork enhances its overall ability to be creative, and if so, how. This is a more complex research challenge, involving new tool development that leverages GenAI for innovative purposes, with functionality that can be invoked (e.g., mundane activities, rapid prototyping) or could be more autonomous in nature (e.g., moderation, contextual information).  For the latter case, less data is available concerning such capabilities or at the very least the data that is available in the GenAI must be harnessed by novel tools in a creative (pun intended) manner.  How individuals respond to such functionality and whether they are comfortable working with or under the direction of GenAI will equally need to be studied.



\subsection{The Product}
Developers build products and, as the Product tetrad reveals (Table \ref{tab:product_tetrad}), many potential impacts exist on the product and they are not all positive. On the one hand, by virtue of GenAI identifying best practices (e.g., UI patterns, architectural models) and  novel ideas, the quality of the code underlying a product could increase and the product could provide tailored user experiences, leading to delighted customers. Additionally, writing less code might lead to developers being less fatigued and thus delivering higher-quality code. On the other hand, developers may well trust results from GenAI too much and not inspect them as they should, leading to poorer quality.  Longer term,  GenAI usage may 
lead to homogenized products.
Anchoring the research on the impact on customers is key since they are the ultimate recipients of the product.  At the same time, understanding the entire ecosystem of all software being produced and whether it as a whole becomes more creative (diverse, innovative) is of equal importance. This, perhaps, could be studied on a domain-by-domain basis (e.g., games, finance).


\subsection{Unintended Consequences}
In focusing on the \say{obsolesces} and \say{reverses into} impacts across all four tetrads (Tables \ref{tab:person_tetrad}-\ref{tab:pres_tetrad}), there are unintended consequences to the person, product, process, and Press from adopting GenAI for software development. Some examples include the potential for a reduction in mentorship from experienced colleagues when GenAI is the sparring partner, biases introduced more readily due to the reliance on GenAI being the creative, and a loss of expertise in software developers due to the reliance on GenAI for software development, with a resulting loss of creativity. These unintended consequences may take time to play out so it may not be possible to research these unintended consequences immediately. At the same time, it is probably essential to start thinking now about how to assess whether such effects may be taking place in future, so that appropriate metrics can be developed and preliminary assessment benchmarks can be performed against which future effects can be measured. 


\subsection{Societal Impact}
Products and people sit in society so it is important to understand how society is impacted by the use of GenAI for creativity in software development. While some of the impacts in the tetrads allude to potential societal impact (e.g., making the impossible possible in Table \ref{tab:product_tetrad}), this is not covered in the 4Ps framework. 
One area to explore is the feelings of society on adopting new products created through the use of GenAI rather than human creativity while another could look at the impact on the software profession. Maybe developers go the way of fellow artisans such as bakers and craft beer makers with a premium attached to unique, hand-made products or they go the way of gas-lighters and no longer relevant to society. Such research is likely to require multi-disciplinary experts such as psychologists, sociologists, and economists alongside ethnographic specialists. We feel that it is particularly important to monitor this aspect of the current wave of GenAI. While the complete demise of the software developer profession may not come to fruition, software is as likely as other disciplines to undergo significant transformations -- at its own hand. Whether or not this is preferable from a societal point of view is the domain of policy and governance; to do so effectively requires a thorough understanding of what is transpiring, which in turn requires carefully constructed longitudinal studies.

\subsection{Human Aspects}
Finally, we must not forget the human behind all of this. Software is built by humans and even the most fantastical scenarios still recognize that humans will be required \cite{freethink_2023}. It is unknown how individual software developers feel about the use of GenAI for creativity and how it impacts their sense of value and purpose, job satisfaction, emotional well-being, and so on. Going beyond the feelings, exploring the actions is also  needed. Are developers actively engaging with GenAI for creative purposes, are they ignoring it and preferring to rely on their own mental capabilities, or, as noted in Table \ref{tab:pres_tetrad} and common in workplace automation initiatives \cite{mueller_breaking_2021}, are they perhaps sabotaging GenAI use? Similar to societal impact, this theme would likely require cross-disciplinary support. 

\section{Conclusion}
Creativity and GenAI are inextricably linked when it comes to software development. No matter how GenAI evolves, people are already using it to help in their creative design work, in their problem-solving when facing coding issues, and more. They look for it to get them started rather than staring at a blank sheet of paper, to nudge them out of fixations, to generate potential root causes for a bug, and so on. In this paper, we examined this relationship by using McLuhan's tetrad to hypothesize how GenAI might impact four different components of creativity (the person, the product, the process, and the Press (environment)). We used these hypotheses to derive a research agenda that can be followed for both understanding the phenomenon as it plays out and for helping to shape it. We feel the latter is particularly important, because with the many potential impacts captured in the tetrads, it becomes possible to design a future with eyes fully open to the potential positive and negative impacts that GenAI may have on many facets of creativity, and ultimately society. This is something behooven of all researchers as society faces uncertain times ahead due to the disruptive influence of GenAI across many aspects of society (including software development). We encourage researchers to consider the potential impacts outlined in this paper, to augment them further by adding their own thoughts and projections, and to join us in researching this phenomenon so we can have a constructive collective voice.



\section*{Use of Generative AI}

GenAI was neither harmed nor used in producing this paper.

\bibliographystyle{ACM-Reference-Format}
\bibliography{refs}

\end{document}